\documentclass[11pt,a4paper]{article}
\usepackage[]{acl2021}
\usepackage{times}
\usepackage{latexsym}

\usepackage{microtype}

\aclfinalcopy 
\def\aclpaperid{2882} 

\usepackage{microtype}
\usepackage{graphicx}
\usepackage{subfigure}
\usepackage{booktabs} 
\usepackage{tikz}
\usetikzlibrary{bayesnet}
\usetikzlibrary{arrows}
\usepackage{color}
\usepackage[toc,page]{appendix}
\usetikzlibrary{backgrounds}
\usepackage{amsfonts}
\usepackage{nicefrac}       
\usepackage{microtype}      
\usepackage{mathtools}
\usepackage{wrapfig}
\usepackage{caption}
\usepackage{amssymb}
\usepackage{amsmath}
\usepackage{float}

\newcommand{\zo}{{\mathbf{z}_o}}
\newcommand{\yo}{{\mathbf{y}_o}}
\newcommand{\zl}{{\mathbf{z}_l}}
\newcommand{\Yt}{{\mathbf{Y}_t}}
\newcommand{\X}{{\mathbf{X}}}
\newcommand{\yl}{{\mathbf{y}_l}}

\title{Learning Robust Latent Representations for Controllable Speech Synthesis}
\author{Shakti Kumar\textsuperscript{*†} \\\And
  Jithin Pradeep\textsuperscript{*}\\\And
    Hussain Zaidi\textsuperscript{*} \\\AND
    {\normalfont \textsuperscript{*}Vanguard Center for Analytics and Insights} \\
    {\normalfont \textsuperscript{†}Computer Science, University of Toronto} \\ \\
    {\normalfont{\{shakti\_kumar, jithin\_pradeep, hussain\_zaidi\}@vanguard.com}}
    }
\date{}
\begin{document}
\maketitle

\begin{abstract}
State-of-the-art Variational Auto-Encoders (VAEs) for learning disentangled latent representations give impressive results in discovering features like pitch, pause duration, and accent in speech data, leading to highly controllable text-to-speech (TTS) synthesis. However, these LSTM-based VAEs fail to learn latent clusters of speaker attributes when trained on either limited or noisy datasets. Further, different latent variables start encoding the same features, limiting the control and expressiveness during speech synthesis. To resolve these issues, we propose RTI-VAE (Reordered Transformer with Information reduction VAE) where we minimize the mutual information between different latent variables and devise a modified Transformer architecture with layer reordering to learn controllable latent representations in speech data. We show that RTI-VAE reduces the cluster overlap of speaker attributes by at least 30\% over LSTM-VAE and by at least 7\% over vanilla Transformer-VAE.
\end{abstract}

\section{Introduction}
Learning disentangled latent representations in speech is an active area of research \citep{DBLP:conf/nips/HsuZG17, DBLP:conf/interspeech/ChouYLL18, Cotatron} with applications in controlling the style (for example, pitch, pause duration, and accent) of synthesized speech.
Recurrent architectures like Long Short Term Memory (LSTM) \citep{LSTM} networks in Variational Autoencoders (VAE) have been state-of-the-art in discovering disentangled latent representations in speech \citep{DBLP:conf/icml/WangSZRBSXJRS18, DBLP:conf/nips/JiaZWWSRCNPLW18, skerryryan2018endtoend} as well as sequential data more generally.
For example \citet{DBLP:conf/icml/LiM18a} attempt to disentangle global and local features of video/speech in different latent variables. \citet{main_conditional_tts} disentangled different dimensions of the latent variables to discover meaningful representations and hence proposed a speech synthesis model with controllable pitch, pause duration, and speed.

These papers as well as several others \citep{DBLP:journals/corr/ChungKDGCB15, 8683561, DBLP:conf/icassp/LeglaiveAGH20, Hono2020, 9053520} 
make one limiting assumption--- the availability of hundreds of hours of speech data for training deep learning networks.
As we show in our experiments, state-of-the-art VAEs fail to learn meaningful separation of speaking styles in speech data when presented with small datasets. In addition, different latent variables learned by the VAE are no longer uncorrelated. Both these shortcomings lead to poor control of speaking styles during synthesis.

While LSTMs are state-of-the-art in learning latent variables in speech, Transformers have been used for understanding latent representations for text completion \citep{TCVAE} and Transformer-based VAEs were used in \citet{9054554} to model independent style attributes in music generation.

Inspired by these limitations of LSTM-based VAEs and the promise of more "attentive" networks, we modify the loss function of the state-of-the-art VAEs \citep{main_conditional_tts} by explicitly minimizing the  mutual  information between latent variables, thereby penalizing common learned features between different representations. We then modify Transformer architecture for learning robust disentangled latent representations of speech from limited and noisy data. We show that our proposed architecture-- RTI-VAE (Reordered Transformer with Information reduction VAE) discovers compact stable latent representations of speaker attributes even on datasets as small as 4 hours of total speech samples while state-of-the-art fails. Our proposed VAE outperforms LSTM and vanilla Transformers even on challenging dataset like Common Voice which has considerable background noise, low recording quality and large number of speakers with the same style or accent.
To summarize, following are the main contributions of our work, %
\begin{enumerate}
    \item Formulate a modified VAE loss function for speech data and a novel Transformer-based VAE for learning uncorrelated latent variables, thereby allowing more precise control over synthesis compared to the existing state-of-the-art.
    \item Show that our latent clusters of speaking styles are better separated than existing LSTM and vanilla Transformer based VAEs on noisy and small datasets.
    \item Show that the our modified Transformer architecture allows a faster convergence of the variational lower bound compared to both vanilla Transformer and LSTM based VAEs.
\end{enumerate}

\section{Related Work}
Multiple previous work have targeted this problem of learning latent representations for sequential data like speech \citep{DBLP:conf/icml/WangSZRBSXJRS18, DBLP:conf/nips/JiaZWWSRCNPLW18, skerryryan2018endtoend}. As discussed, the main advantage of learning such representations is that it allows creating diverse examples during reconstruction by manipulating the encoded latent variable. In \citet{DBLP:conf/icml/LiM18a} the authors propose two sets of latents which learn global features like the generated sequence contents and local dynamic features such as pitch, speed etc. However, a limitation of this approach is the lack of interpretability of the learnt dimensions--- it is known that the different dimensions of the latent variables are learning some features but there is little to no visibility into what those actual features are.


Modifying Text-to-Speech systems by introducing additional encoders has been a standard way to discover meaningful representations. \citet{DBLP:conf/icassp/ZhangPHL19} build on top of Tacotron-2 \citep{tacotron-2} architecture and use Gaussians to model their latent variables. An improved version can be seen in \citet{main_conditional_tts} where a hierarchical latent with mixture of Gaussians is used.
\citet{8683561} propose adversarial training to further improve latent variables and the features discovered by disentangling the background noise and reverberation along with speaker identity from the recording conditions.



While all these prior work aim to discover latent representations, there is a lot of room for improving those representations especially in cases where we have very limited hours of speech dataset. As we show in our experiments, in the absence of explicit restrictions on the training objective these VAEs easily collapse when presented with smaller datasets. Thus we focus on improving the representations, specifically latent clusters of speaker attributes, in cases of extremely limited datasets. Our contributions, however are not limited to smaller datasets and we see similar improved performance on larger and noisy datasets too.


\section{Background}
\begin{figure}
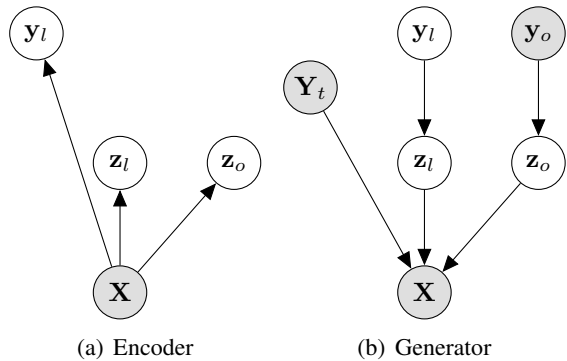

\begin{center}
    \subfigure[Encoder]{
          \tikz{
         \node[obs] (x) {$\X$};%
         \node[latent,above=of x] (zl) {$\zl$}; %
         \node[latent,above=of x,xshift=1.5cm,fill] (zo) {$\zo$}; %
         \node[latent,above=of zl,xshift=-1.1cm] (yl) {$\yl$}; %
         \edge{x}{zl, zo, yl}
         }
    }
    \subfigure[Generator]{
          \tikz{
         \node[obs] (x) {$\X$};%
         \node[obs,above=of x,xshift=-1.5cm,yshift=1cm,fill] (yt) {$\Yt$}; %
         \node[latent,above=of x] (zl) {$\zl$}; %
         \node[latent,above=of x,xshift=1.5cm,fill] (zo) {$\zo$}; %
         \node[obs,above=of zo] (yo) {$\yo$}; %
         \node[latent,above=of zl] (yl) {$\yl$}; %
         \edge{yt,zo,zl} {x}
         \edge{yl} {zl}
         \edge{yo}{zo}
         }
    }
\caption{Graphical model of controllable TTS system. Note that $q(\yl|\X)$ in the Encoder can be approximated in terms of $q(\zl|\X)$, in which case node $\yl$ will have an edge from $\zl$ instead of $\X$ as done in \citet{main_conditional_tts}.}
\label{fig:graphical_model}
\end{center}
\end{figure}

Controllable text-to-speech (TTS) VAE-based systems like in \citet{main_conditional_tts} take an input text sequence $\Yt$ and an optional observed categorical label $\yo$ (e.g., speaker identity or accent) as input and learn to synthesize a sequence, usually mel-spectrogram frames $\X$ as output. Additional latent variables $\zo$ and $\zl$ can be introduced to discover meaningful representations during this process. Here $\zo$ is a continuous latent learnt on top of shown labels $\yo$, hence $\zo$ captures the variation in features correlated with the speaker attribute $\yo$. $\zl$ is a completely unsupervised continuous variable learnt on top of standard Expectation-Maximization style latent mixture components $\yl$. This graphical model is depicted in Figure \ref{fig:graphical_model}. The objective function for learning such model, i.e. synthesizing sequence $\X$ given $\Yt$ and $\yo$, can be formulated as the variational lower bound\footnote{Complete derivation is given in the Appendix.},

\begin{flalign}
        log\;p(&\X |\Yt, \yo)
        \geq \;log\;p(\X|\Yt, \widetilde{\zo}, \widetilde{\zl}) \nonumber \\
        &-\sum_{\yl=1}^{K} q(\yl | \X)D_{KL}[\;q(\zl | \X)\;||\;p(\zl | \yl)\;] \nonumber \\
        &- D_{KL}[\;q(\yl|\X)\;||\;p(\yl)\;] \nonumber\\
        &- D_{KL}[q(\zo|\X)\;||\;p(\zo | \yo)\;] \nonumber\\
        &= -L_{mel} -L_{KL} \nonumber
\end{flalign}

where $L_{mel} = -log\;p(\X|\Yt, \widetilde{\zo}, \widetilde{\zl})$ and $L_{KL}$ refers to the remaining terms. Here $\widetilde{\zo}, \widetilde{\zl}$ are sampled points and are reparameterized \citep{gaussian_reparam} as $\widetilde{\zo} = \hat{\mu_o} + \hat{\sigma_o} \odot \epsilon_o$ and $\widetilde{\zl} = \hat{\mu_l} + \hat{\sigma_l} \odot \epsilon_l$ with $\hat{\mu_o}, \hat{\mu_l}, \hat{\sigma_o}, \hat{\sigma_l}$ as the mean and standard deviation of the posterior distributions $q(\zo|\X)$ and $q(\zl|\X)$ respectively and with auxiliary noise variable $\epsilon_o, \epsilon_l \sim \mathcal{N}(0, \textit{I})$. Following \citet{beta_vae} the loss $L$ can be written in a more general form as,
\begin{flalign}
    &L = L_{mel} + \beta L_{KL} \label{loss_without_MI}
\end{flalign}
with $\beta$ balancing the relative weighing between the latent channels and reconstruction accuracy. Here $L_{mel}$ is the mel loss which controls the quality of the mel-spectrograms produced and $L_{KL}$ refers to the total KL Loss controlling the features learnt in latent variables.

This VAE can be used in the Tacotron-2 architecture \citep{main_conditional_tts} as shown in Figure \ref{fig:tacotron_architecture} to learn the text to mel-spectrogram mapping and the latent features controlled by $L_{KL}.$

\section{Methodology}
We now describe the two main components, 1) Minimizing mutual information and 2) Layer reordering in our proposed RTI-VAE architecture.

\begin{figure*}
\begin{center}
    \subfigure[Tacotron Architecture]{\includegraphics[scale=0.4]{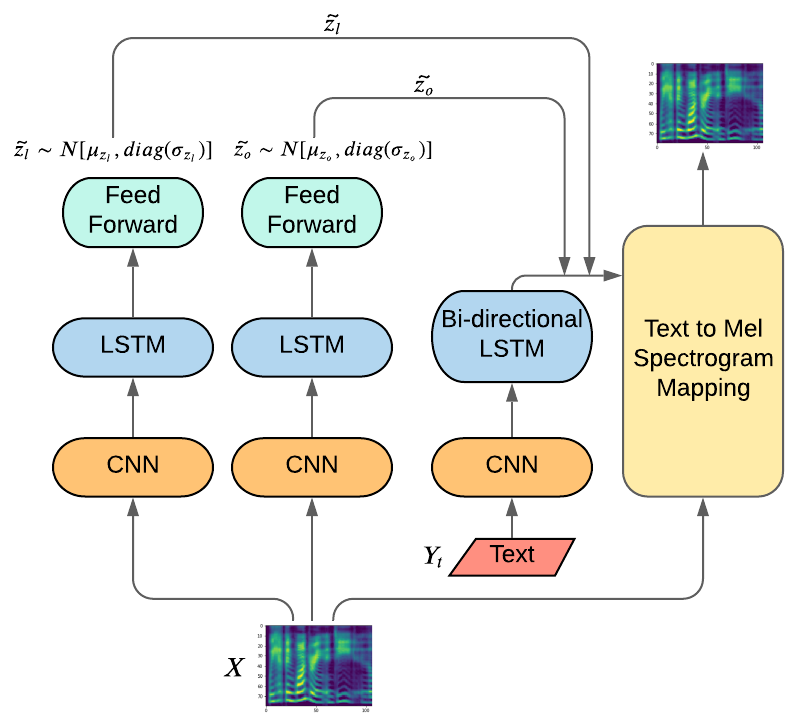}
    \label{fig:tacotron_architecture}
    }
    \hspace{20pt}
    \subfigure[Proposed Encoder]{
          \tikz{
         \node[obs] (x) {$\X$};%
         \node[latent,above=of x] (zl) {$\zl$}; %
         \node[latent,above=of x,xshift=1.5cm,fill] (zo) {$\zo$}; %
         \node[latent,above=of zl,xshift=-1.1cm] (yl) {$\yl$}; %
         \node[obs,above=of zo] (yop) {$y_{o}$};
         \edge{x}{zl, zo, yl}
         \edge{zl}{yop}
         \draw[above](zl) edge node{$q_\psi$} (yop)
         }
    \label{fig:proposed_encoder}
    }
    \hspace{10pt}
    \subfigure[Original versus Proposed Transformer]{
        \includegraphics[scale=0.35]{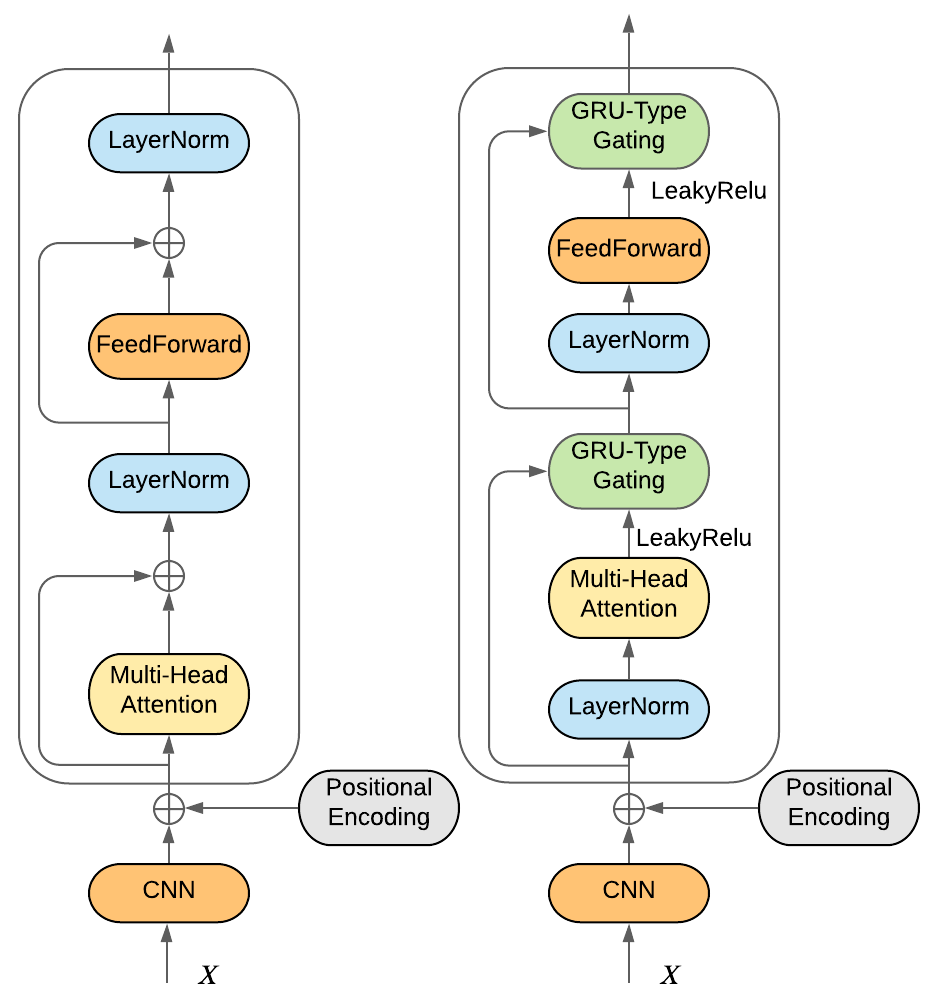}
    }
    \caption{
    \textbf{Left}: The Tacotron-2 architecture. VAE consists of two left parts where LSTMs predict mean $\mu$ and variance $\sigma^2$ of multivariate Gaussians $\mathcal{N}(\mu_{\zl}, diag(\sigma_{\zl}^2)), \mathcal{N}(\mu_{\zo}, diag(\sigma_{\zo}^2))$. $\widetilde{\zl}, \widetilde{\zo}$ from this distribution are sampled and concatenated to the text encoding to conditionally learn the text to mel-spectrogram mapping.
    \textbf{Center}: Proposed encoder with the network $q_{\psi}$. The generator stays the same as in Figure \ref{fig:graphical_model}.
    \textbf{Right}: The original and the proposed Transformers replace the LSTMs shown in the VAE of Tacotron-2 architecture.}
    \label{fig:model}
\end{center}
\end{figure*}
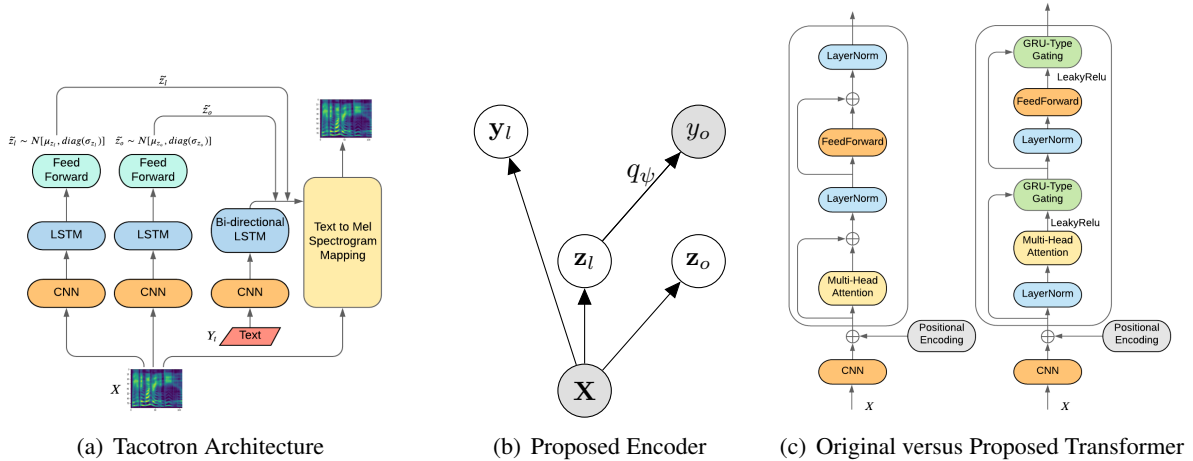{}
\subsection{Minimizing Mutual Information} \label{sec:minimizing_mutual_info}
The latent $\zl$ in Figure \ref{fig:graphical_model} is unsupervised while the latent $\zo$ learns features correlated with the shown label $\yo$. Our experiments showed that both $\zl, \zo$ can end up encoding the same set of features, which leads to poor control in synthesizing speech. An intuition into why this happens lies in the fact that $\zl$ is an unsupervised variable and it can discover any feature hidden in the input speech sequence. There is no term in the loss function \eqref{loss_without_MI} which prevents the features of $\zl$ from being correlated with the observed labels $\yo$ \citep{min_latent}.

This can be resolved by minimizing the mutual information $I$ between latents $\zo$ (equivalently $\yo$) and $\zl$. We can formulate this as,
\begin{equation*}
\begin{aligned}
        & min\;I(\yo;\zl)
        \triangleq max\;H(\yo|\zl) \\
        &= min\;\int_{\zl}\int_{\yo}p(\zl)\;p(\yo|\zl)\;log\;p(\yo|\zl)d\yo d\zl \\
        &= min\;\int_{\X}\int_{\zl}\int_{\yo}{\begin{multlined}p(\X)\;p(\zl|\X)\;p(\yo|\zl)\\log\;p(\yo|\zl)\;d\yo\;d\zl\;d\X\end{multlined}}
\end{aligned}
\end{equation*}
Since integral over $\zl$ is intractable, we replace $p(\zl|\X)$ with an approximate posterior $q(\zl|\X)$. Further, since the true distribution $p(\yo|\zl)$ is unknown, we approximate it by introducing a new network
$q_{\psi}(\yo|\zl)$ leading to $min\;I(\yo;\zl)$
\begin{equation}
\begin{aligned}
    & \approx min\int_{\X}\int_{\zl}\int_{\yo}{\begin{multlined}
    p(\X)\;q(\zl|\X)\;q_{\psi}(\yo|\zl) \\log\;q_{\psi}(\yo|\zl)d\yo\;d\zl\;d\X
    \end{multlined}} \\
    &= min\;E_{D(\X)q(\zl|\X)}\left[\int_{\yo}{\begin{multlined}q_{\psi}(\yo|\zl) \\log\;q_{\psi}(\yo|\zl)\;d\yo\end{multlined}} \right]\\
    &\approx min\;\frac{1}{N}\sum_{a}
    \left[{\begin{multlined}
    q_{\psi}(\yo=a|\zl')\\log\;q_{\psi}(\yo=a|\zl')
    \end{multlined}}\right]
\label{first_mutual_terms}
\end{aligned}
\end{equation}
where $\zl' \sim q(\zl|\X)$, $a\in{\{0,1,2...A\}}$, $A$ is total number of unique classes of $\yo$, $N$ is the number of samples used for Monte Carlo estimates, and $D(\X)$ is the underlying distribution of the input points $\X$. Our proposed encoder is depicted in Figure \ref{fig:proposed_encoder}. Since we are using $q_{\psi}$ to make predictions for $\yo$, this network needs to be learnt itself. Hence we need to subtract an additional $q_{\psi}(\yo|\zl')$ from the loss function. With $N=1$ our proposed term is,
\begin{flalign}
    L_{MI}  &= \sum_{a}q_{\psi}(\yo=a|\zl')log\;q_{\psi}(\yo=a|\zl')  \nonumber\\
            &- q_{\psi}(\yo|\zl')
\label{total_mutual_term}
\end{flalign}

Combining equations \eqref{loss_without_MI} and \eqref{total_mutual_term}, the total loss function in our proposed model is,
\begin{flalign}
        L_{total} &= L_{mel} + \beta L_{KL} + \gamma L_{MI} \label{total_loss} \\
                  &= L_{mel} + L_{cond} \nonumber
\end{flalign}

To summarize, $L_{mel}$ controls the quality of the mel-spectrogram produced during decoding, $L_{KL}$ controls the features learnt in the latent variables $\zl, \zo$ and $L_{MI}$ makes sure that $\zl, \zo$ encode different features. We will be referring to $L_{mel}$ as the reconstruction or mel loss, $L_{KL}$ as the KL loss and $L_{cond} = \beta L_{KL} + \gamma L_{MI}$ as the conditional loss respectively throughout this paper.

\subsection{Layer Reordering in Transformer} \label{sec:transformer_based_vae}
Introducing the above loss helps disentangle the learning of $\zo$ and $\zl$, but there is another problem that remains. Our experiments on MAILABS and Common Voice data, discussed in section \ref{sec:cluster_quality}, indicated that clusters of $\zo$ corresponding to different shown labels $\yo$ start sharing regions in the latent space.
Hence for any given label $\yo$ the sampled $\hat{\zo}\sim p(\zo|\yo)$ may or may not belong to the style which $\yo$ denotes. This leads to speech samples where the style correlated with the shown attribute $\yo$ is not under control while sampling from the priors.

\begin{table*}
\centering
\begin{tabular}{l*{5}{c}r}
\hline \textbf{$d$}   & \textbf{Feature}  &$\mu_{\zl, d}-3\sigma_{\zl, d}$ & $\mu_{\zl, d}$& $\mu_{\zl, d}+3\sigma_{\zl, d}$ \\ \hline
0 & Speaking Rate (sec)          & $3.0\pm0.2$       & $3.7\pm0.3$       & $4.4\pm0.3$ \\
1 & $F_0$ (Hz)              & $240.5\pm12.57$    & $211.4\pm15.66$    & $184\pm10.43$ \\
2 & Pause Duration (msec)   & $70\pm3.40$       & $79\pm3.30$       & $91\pm3.50$ \\
\hline
\end{tabular}
\caption{Length of the mel-spectrogram synthesized and pause durations increase while pitch decreases with increasing $d$th dimension of $\zl$ from its marginal prior mean in RTI-VAE.}
\label{tab:features_learnt}
\end{table*}

\begin{figure*}[ht!]
\begin{center}
    \includegraphics[width=\textwidth]{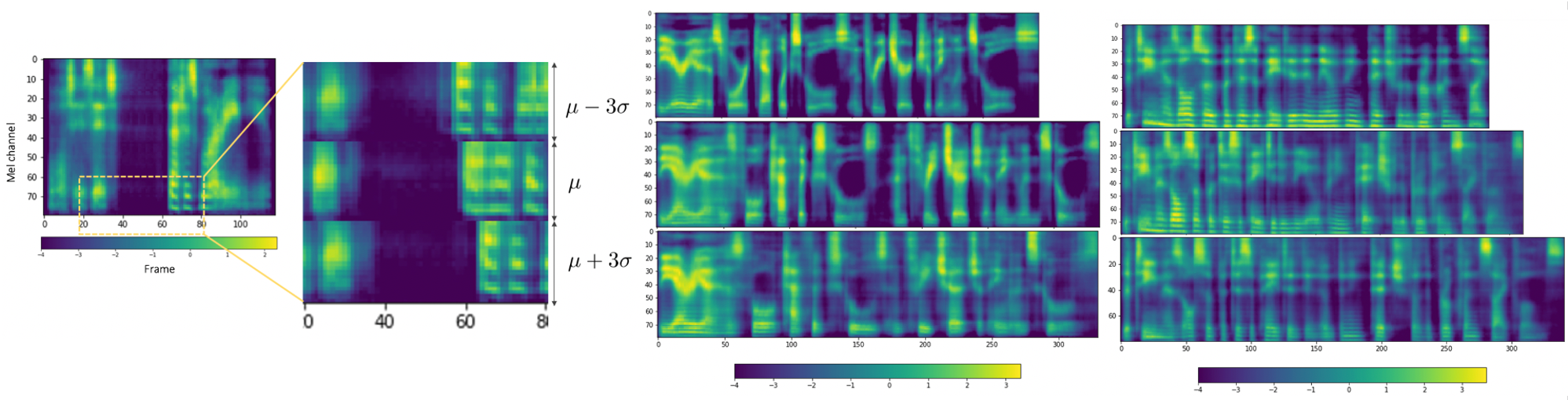}
    \caption{\textbf{Left}: Synthesized mel spectrogram for \textit{"What is it, that is worrying you today?"} The stack of 3 mel spectrograms on the right are zoomed areas from frames 20 to 80 for each of their original mel-spectrogram. It can be seen that the pause duration denoted by the dark region increases as you synthesize the same text moving from $\mu_i-3\sigma_i$ to $\mu_i+3\sigma_i$.
    \textbf{Center}: Three mel-spectrograms synthesized for the text \textit{"The area has four catholic schools and three church of England schools"}, corresponding to three random sampling of $\widetilde{\zo}, \widetilde{\zl}$ from their posteriors. First synthesis is considerably shorter than the second and third. Notice the different positions of voids between frames 50 and 100, and at frame 150 in the third spectrogram being considerably different.
    \textbf{Right}: Mel-spectrograms synthesized for the text \textit{"The team has also participated in the opening pitch of the Brooklyn Cyclones"}. The third spectrogram shows smooth areas in the higher mel channels compared to the second and the first. These random latent sampling affects intonation and spectrogram texture.
    }
    \label{fig:pause_duration_sampling_latents}
\end{center}
\end{figure*}{}

We tackle this problem by replacing LSTMs with Transformers.
We expected that the ability of Transformers to attend to specific frames of interest where features could be localized or have a higher expression density, with a higher weight in the input speech sequence should bring down the dataset volume required for convergence by a considerable amount. Hence the lower bound on dataset size needed for modelling non overlapping clusters of $\zo$ should be smaller while still keeping the sampled style under control. This should also accelerate the separation between latent clusters for larger datasets. Our experiments with vanilla Transformer-based VAEs confirm our predictions.

We next drew some inspiration from \citet{RL} and modified the Transformer encoder. This was an attempt at changing the learning paradigm--- instead of directly learning to translate $\Yt$ to $\X$ in different $\yo$ styles, we first learn to synthesize a general representation for all $\X$, and then learn specific deviations of each style $\yo$ from this general representation. For example, instead of learning directly to speak in different accents first we learn to speak, and then we learn the subtleties of different accents. Our hypothesis was that learning different $\yo$ styles should be a lot faster if a common understanding of all $\X$ in the dataset is gained first. The accent specific speech frames $\X$ (or style specific as per $\yo$) should just be a slight deviation from this common representation.

Our proposed architecture is shown in Figure \ref{fig:model}c where we switch the order of \texttt{LayerNorm} forming a direct connection between the input and the output.
Due to this layer reordering if we make sure that all the modules \texttt{MHA, LayerNorm, FeedForward} are initialized with their expectation near 0, a direct path is formed early in training allowing a general representation of speech to be learnt independent of the shown labels $\yo$. Now as training progresses and these modules warm up, the accent or $\yo$ specific features will be learnt by conditioning the encoder.

We also introduce GRU-type gating \citep{GRU} to stabilize learning by minimizing the maximum gradient norms produced, and apply a small nonlinearity via $LeakyRelu$ at the outputs of the \texttt{MHA} and \texttt{FeedForward} modules to balance the observed trade-off between frequent gradient updates and maximum gradient norm\footnote{The specific choice of $LeakyRelu$ is discussed in the Appendix.}.


\section{Experiments}
We refer to our proposed VAE with modifications from sections \ref{sec:minimizing_mutual_info} ($L_{MI}$ term) and \ref{sec:transformer_based_vae} as RTI-VAE, the vanilla Transformer with $L_{MI}$ term as Transformer-VAE and the LSTM based state-of-the-art Tacotron-2 without $L_{MI}$ term \citep{main_conditional_tts} as LSTM-VAE. We trained each model on two datasets--- 1) MAILABS \citep{mailabs} with a total 35hrs of UK and 39hrs of US speech in studio quality recorded by 4 professional speakers, 2) Common Voice \citep{CommonVoice} with 4hrs of UK and 19hrs of US speech crowd-sourced from 477 volunteers with varying background noise, microphone qualities and other recording conditions.
The input feature $\X$ were mel-scale spectrograms,
the label $\yo$ was set to be 0 for all $\X$ belonging to US and 1 for all UK. Dimension of $\zo$ and $\zl$ were picked to be 2 and 3 respectively and $K$ = 3 for all experiments
\footnote{Other hyperparameters of our VAE and training details of Tacotron-2 are given in the Appendix.}.

\subsection{Features Learnt}
Before we demonstrate our latent cluster improvements over Transformer-VAE and LSTM-VAE, we show that RTI-VAE does learn important latent features in speech. Our experiments (focused on learning the speaking rate, the fundamental frequency $F_0$, and the pause duration) are summarized in Table \ref{tab:features_learnt}. $\mu_{\zl, d}$ and $\sigma_{\zl, d}$ are the $dth$ dimension mean and standard deviations of the marginal prior $p(\zl) = \sum_k{p(\zl|\yl=k)p(\yl=k)}$. All other dimensions of $\zl$ are kept fixed at their own marginal priors while analyzing $d$th dimension.

For demonstrating control on speaking rate, we did 25 different synthesis for the text \textit{"We had been wandering, indeed, in the leafless shrubbery an hour in the morning"}. It can be seen from Table \ref{tab:features_learnt} that the length of the synthesized mel-spectrogram increases as the value of $\zl$ dimension 0 increases.

\begin{figure*}[ht!]
    \centering
    \includegraphics[width=\textwidth]{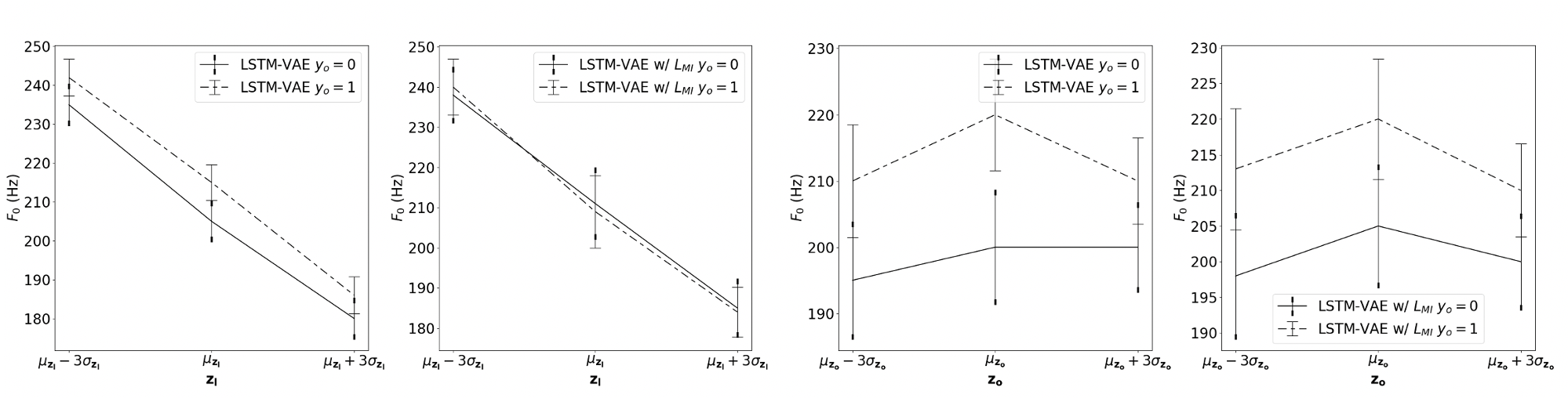}
\caption{In LSTM-VAE $F_0$ encoded by $\zl$ is significantly different for $\yo=0,1$ showing that $y_o$ specific information is encoded by $\zl$. However this difference is no longer significant once we include our proposed $L_{MI}$ terms in LSTM-VAE w/ $L_{MI}$ experiment. $\zo$ keeps showing different values of $F_0$ for $y_o=0,1$ in both LSTM-VAE and LSTM-VAE w/ $L_{MI}$ experiments demonstrating learnt features which are conditional on $y_o$.
}
\label{fig:f0_trends_lmi}
\end{figure*}{}

\begin{figure*}[ht!]
    \centering
    \includegraphics[scale=0.6]{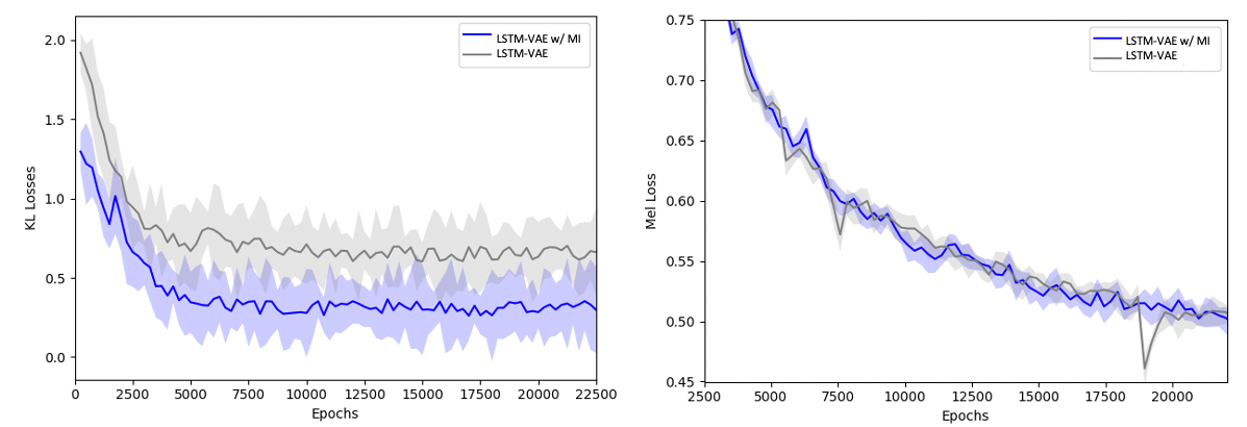}
\caption{\textbf{Left}: Test $L_{KL}$ versus epochs. Including $L_{MI}$ in loss function decreases $L_{KL}$ pointing to improved latent variables. \textbf{Right}: Test $L_{mel}$ versus epoch. The $L_{mel}$ remains the same even upon including $L_{MI}$ demonstrating our proposed $L_{MI}$ does not hurt the synthesized mel-spectrogram quality.
}
\label{fig:importance_of_lmi}
\end{figure*}{}

Next, we synthesized 25 texts, with 10 samples for each text to show control on pause duration and pitch (or the fundamental frequency $F_0$). For pause duration experiments each text contained at least one comma and we measured the maximum period of intermediate silence for each synthesis. To calculate $F_0$ we used the YIN algorithm \citep{patrice_guyot_yin}. In Table \ref{tab:features_learnt} it can be seen that the pause duration increases and $F_0$ decreases with increasing values of 2nd and 1st dimensions of $\zl$, respectively.

Furthermore the sampled variables $\widetilde{\zo}, \widetilde{\zl}$ from their respective posterior distributions $q(\zo|\X), q(\zl|\X)$ in $L_{mel}$ gives the effect of different intonations with different speakers every time we synthesize a given text $\Yt$. We demonstrate concrete examples in Figure \ref{fig:pause_duration_sampling_latents}.

\begin{table*}[ht!]
    \centering
    \begin{tabular}{l*{7}{c}r}
        \hline
            & \multicolumn{2}{c}{\textbf{4hrs US+4hrs UK}}
        & \multicolumn{2}{c}{\textbf{20hrs US+20hrs UK}}
        & \multicolumn{2}{c}{\textbf{39hrs US+35hrs UK}} \\
        \textbf{Model}   &   \textbf{DI}  &   \textbf{DBI}  &   \textbf{DI}  &   \textbf{DBI} &   \textbf{DI}  &   \textbf{DBI} \\
        \hline
        LSTM-VAE        &   0.55$\pm$0.15  &    2.11$\pm$0.24   &   1.41$\pm$0.21    &   1.60$\pm$0.29    &   2.10$\pm$0.29    &   1.12$\pm$0.24 \\
        Transformer-VAE &   1.22$\pm$0.26   &   0.44$\pm$0.05   &   2.24$\pm$0.05    &   0.30$\pm$0.15    &   2.48$\pm$0.23    &   0.27$\pm$0.09 \\
        RTI-VAE    &   \textbf{1.85$\pm$0.59}   &   \textbf{0.35$\pm$0.07}   &   \textbf{2.33$\pm$0.21}    &   \textbf{0.29$\pm$0.10}    &   \textbf{2.80$\pm$0.26}    &   \textbf{0.26$\pm$0.07} \\
    \hline
    \end{tabular}
    \caption{RTI-VAE consistently increases DI and reduces DBI for different sizes of MAILABS dataset and performs at least 3\% better (DBI for 20hrs US+20hrs UK) on MAILABS dataset compared to all existing architectures.
}
    \label{tab:di_dbi_mailabs}
\end{table*}
\begin{table*}[ht!]
    \centering
    \begin{tabular}{l*{7}{c}r}
        \hline
            & \multicolumn{2}{c}{\textbf{4hrs US+4hrs UK}}
        & \multicolumn{2}{c}{\textbf{10hrs US+4hrs UK}}
        & \multicolumn{2}{c}{\textbf{19hrs US+4hrs UK}} \\
        \textbf{Model}   &   \textbf{DI}  &   \textbf{DBI}  &   \textbf{DI}  &   \textbf{DBI} &   \textbf{DI}  &   \textbf{DBI} \\
        \hline
        LSTM-VAE        &   0.98$\pm$0.17   &   83.18$\pm$13.66   &   0.85$\pm$0.23    &   85.53$\pm$15.10    &   0.80$\pm$0.30    &   98.20$\pm$24.68 \\
        Transformer-VAE &   0.99$\pm$0.15   &   0.19$\pm$0.01   &   0.98$\pm$0.22    &   0.18$\pm$0.18    &   0.94$\pm$0.29    &   0.17$\pm$0.30 \\
        RTI-VAE    &   \textbf{1.03$\pm$0.40}   &   \textbf{0.15$\pm$0.005}   &   \textbf{0.99$\pm$0.20}    &   \textbf{0.16$\pm$0.04}    &   \textbf{0.99$\pm$0.25}    &   \textbf{0.16$\pm$0.05} \\    \hline
    \end{tabular}
    \caption{RTI-VAE performs at least 4\% better (DI for 4hrs US+4hrs UK Common Voice compared to Transformer-VAE) on all sizes of noisy Common Voice dataset than all existing LSTM and Transformer-VAE architectures.
}
    \label{tab:di_dbi_cv}
\end{table*}
\begin{table*}[ht!]
    \centering
    \begin{tabular}{l*{7}{c}r}
        \hline
            & \multicolumn{3}{c}{\textbf{Overlap on MAILABS}}
        & \multicolumn{3}{c}{\textbf{Overlap on Common Voice}} \\
        \textbf{Model}   &   \textbf{4+4}  &   \textbf{20+20}  &   \textbf{39+35}  &   \textbf{4+4} &   \textbf{10+4}  &   \textbf{19+4} \\
        \hline
        LSTM-VAE        &   30\%   &   11\%   &   0\%    &   92\%    &   94\%    &   96\% \\
        Transformer-VAE &   7\%   &   0\%   &   0\%    &   52\%    &   65\%    &   81\% \\
        RTI-VAE    &   \textbf{0\%}   &   0\%   &   0\%    &   \textbf{47\%}    &   \textbf{56\%}    &   \textbf{65\%} \\    \hline
    \end{tabular}
    \caption{Overlap percentages for datasets of size $M+N$ with $M$ hrs US and $N$ hrs UK speech. RTI-VAE reduces the overlap percentage by 30\% for limited MAILABS dataset and by half for limited Common Voice dataset. The reduction difference for entire Common Voice dataset is 31\% compared to LSTM and 16\% compared to Transformer-VAE.}
    \label{tab:overlap_percentages}
\end{table*}

\subsection{Importance of $L_{MI}$}
Our experiment on MAILABS dataset shows that the latent variable $\zl$ starts encoding $\yo$ specific features in the absence of an explicit $L_{MI}$ term in the total loss, contrary to the expectation that $\zl$ should not encode any $\yo$ style specific information. As shown in Figure \ref{fig:f0_trends_lmi}, $\zl$ shows different values of $F_0$ for classes $\yo=0,1$ in the absence of $L_{MI}$, while $\zo$ continues to show accent specific values for both $\yo$ classes with and without $L_{MI}$ terms. The values in Figure \ref{fig:f0_trends_lmi} are plotted for a synthesis of 25 different texts with 10 samples for each text. We show similar trends for speaking rate in the Appendix.


A consequence of including $L_{MI}$ in the loss function \eqref{total_loss} can also be seen in the test curve of $L_{KL}$. We can see in Figure \ref{fig:importance_of_lmi} that LSTM-VAE w/ MI has a lower value of $L_{KL}$.
Also note that as shown in Figure \ref{fig:importance_of_lmi}, $L_{mel}$ remains the same in both the experiments hence there is an overall decrease in the total loss value. We also observe that the two terms of $L_{MI}$ in equation \eqref{total_mutual_term} are in contention to each other. The first term tries to learn a representation $\zl$ such that it does not have any information about label $\yo$ whereas the second term tries to maximize the probability of predicting label $\yo$ given $\zl$. We verify from our experiments that at convergence $\zl$ acts as a complete random input for estimating $\yo$ with $q_\psi(\yo|\zl)=0.5$ for both $\yo=0,1$.

\begin{figure*}[ht!]
    \centering
    \includegraphics[width=\textwidth]{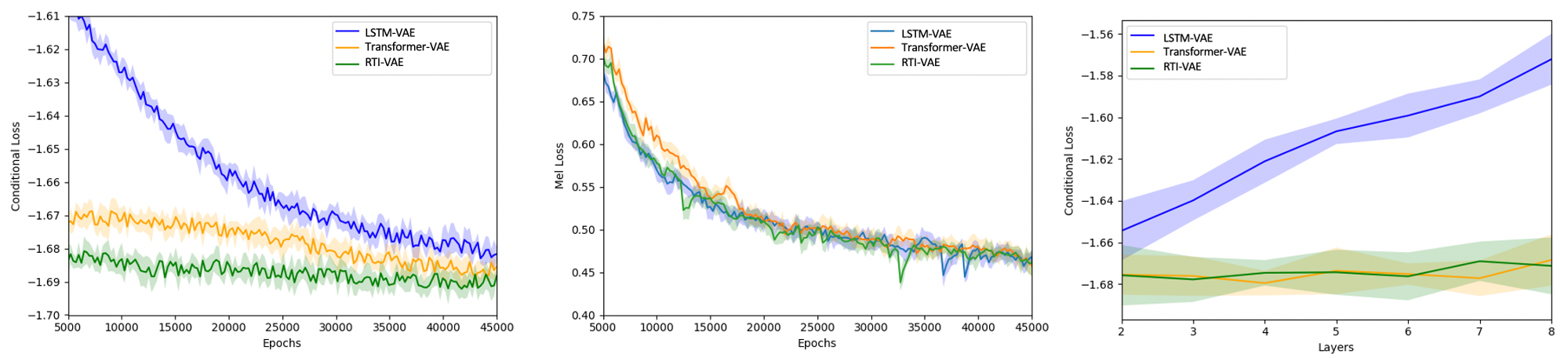}
\caption{
Loss Curves on MAILABS dataset.
\textbf{Left}: Test $L_{cond}$ versus Epochs. RTI-VAE converges faster compared to both Transformer-VAE and LSTM-VAE.
\textbf{Center}: Test $L_{mel}$ versus Epochs. RTI-VAE accelerates $L_{cond}$ without compromising the mel-spectrogram quality or $L_{mel}$.
\textbf{Right}: Test $L_{cond}$ versus model depth. Transformer and RTI-VAE do not overfit to a given dataset with increasing model depth unlike LSTM-VAE.}
\label{fig:loss_curves}
\end{figure*}{}

\subsection{Cluster Quality}\label{sec:cluster_quality}
As discussed in section \ref{sec:transformer_based_vae}, we want clusters of $p(\zo|\yo=0)$ and $p(\zo|\yo=1)$ to be far from each other with no overlaps so that we can control $\yo$ styles during synthesis. Hence we objectively measured the cluster quality with Dunn Index (DI) \citep{DI} and DB Index (DBI) \citep{DBI} where $DI$=$\frac{min_{1\leq i<j\leq n}d(i,j)}{max_{1 \leq k \leq n}d'(k)}$, $DBI$=$\frac{1}{n}\sum_{i=1}^{n}max_{j\neq i}\left( \frac{\sigma_i + \sigma_j}{d(\mu_i, \mu_j)}\right)$, $j,i$ are cluster indices, $d(i,j)$ denotes the distance between the clusters $i$ and $j$, $n$ is the total number of points, $d'(k)$ is the maximal intra-cluster distance and $\mu_i, \sigma_i, \mu_j, \sigma_j$ are the means and standard deviations of the clusters $i, j$ respectively. Thus DI is the ratio of minimal inter-cluster distance to the maximal intra-cluster distance. Similarly, DBI is the ratio of spread in each cluster to the distance between their means.

In Tables \ref{tab:di_dbi_mailabs} and \ref{tab:di_dbi_cv}, we compare the test DI and DBI for different dataset sizes between RTI-VAE, Transformer-VAE and LSTM-VAE. We see that RTI-VAE performs consistently better than Transformer-VAE and LSTM-VAE for both MAILABS and Common Voice dataset. We also observe that as dataset size decreases, the performance gap between our RTI-VAE and LSTM-VAE increases.

In Table \ref{tab:overlap_percentages} we calculate the percentage of overlap between clusters with test points $\hat{\zo}\sim p(\zo|\yo=i)$ marked as overlapping with cluster $p(\zo|\yo=j)$ if they fall within [$\mu_{p(\zo|\yo=j)}+\sigma_{p(\zo|\yo=j)}, \mu_{p(\zo|\yo=j)}-\sigma_{p(\zo|\yo=j)}]$, with $i,j=0,1$. We observe that our RTI-VAE consistently decreases the overlap regions by large margins even on challenging datasets like Common Voice, where more than $90\%$ overlap exists for existing state-of-the-art. As discussed earlier this better separation provides improved control on synthesis and prevents uncontrolled styles when sampling speech from the priors.

\subsection{Loss Curves}
The conditional loss $L_{cond}$ in equation \eqref{total_loss} controls the latent variables being modelled namely $\zl, \zo$ and $\yl$. The trend in Figure \ref{fig:loss_curves} for MAILABS dataset shows that RTI-VAE has an accelerated convergence compared to both Transformer-VAE and LSTM-VAE. It can also be seen in Figure \ref{fig:loss_curves} that $L_{mel}$ remains the same in all the 3 experiments, LSTM-VAE, Transformer-VAE and RTI-VAE. This shows that while our RTI-VAE is successful in lowering $L_{cond}$, it does so without hurting $L_{mel}$ or the synthesized mel-spectrogram quality.

We also observed that for a given dataset size in LSTM-VAE, $L_{cond}$ increases with increasing model depth which points towards inferior latent features. This trend is summarized in Figure \ref{fig:loss_curves} and shows that Transformer-VAE and RTI-VAE do not overfit to a given dataset size with increasing layers.

\section{Conclusion}
In this work we showed that RTI-VAE discovers disentangled latent representations of speech with uncorrelated latent variables allowing better control of speech synthesis. Our layer reordering in Transformers produces notably improved latent clusters of speaker attributes keeping the speaker styles under control on varying dataset sizes with different noise conditions. We can generate mel spectrograms for different text with controllable pitch, pause durations, speaking speed and accent. We also showed that there is a significant boost both in convergence and in the stability of the learnt representations with our proposed method. Going forward we would like to explore the application of RTI-VAE beyond speech, e.g, image captionining with sentiments or text to image rendering with different emotions.

\bibliographystyle{acl_natbib}


\clearpage
\newpage
\section*{Appendix}
\appendix
\section{Variational Lower Bound}\label{supp:variational-lower-bound}
For an input text sequence $Y_t$ and an observed categorical label $y_o$ frames $X$ can be learnt via the joint distribution $log\;p(X, Y_t, y_o)$. Additional latent variables $z_o$ and $z_l$ can be introduced to discover meaningful representations during this process. Here $z_o$ is a continuous latent learnt on top of shown labels $y_o$, hence the features $z_o$ discovers is correlated with what is shown to the model via $y_o$, while $z_l$ is a completely unsupervised continuous variable learnt on top of standard Expectation-Maximization style latent mixture components $y_l$. Note that $y_l$ is a $K$-way categorical discrete variable. The variational lower bound can then be formulated as,


\begin{flalign}
        log\;&p(X|Y_t, y_o) \geq \mathbb{E}_{q(z_o|X)q(z_l|X)q(y_l|X)} \nonumber\\
        &\left[log\frac{p(X|Y_t, z_o, z_l)p(z_o|y_o)p(z_l|y_l)p(y_l)}{q(z_o|X)q(z_l|X)q(y_l|X)}\right] \nonumber\\
        &= \mathbb{E}_{q(z_o|X)q(z_l|X)}[log\;p(X|Y_t, z_o, z_l)] \\
        &- D_{KL}(q(z_o|X)\;||\;p(z_o|y_o)) \nonumber\\
        &- \mathbb{E}_{q(y_l|X)}[D_{KL}(q(z_l|X)\;||\;p(z_l|y_l))] \nonumber\\
        &- D_{KL}(q(y_l|X)\;||\;p(y_l)) \nonumber\\
        &\approx \;log\;p(X|Y_t, \widetilde{z_o}, \widetilde{z_l}) \\
        &-\sum_{y_l=1}^{K} q(y_l | X)D_{KL}[\;q(z_l | X)\;||\;p(z_l | y_l)\;] \\
        &- D_{KL}[\;q(y_l|X)\;||\;p(y_l)\;] \nonumber\\
        &- D_{KL}[q(z_o|X)\;||\;p(z_o | y_o)\;] \nonumber\\
        &= -L_{mel} -L_{KL} \nonumber
\end{flalign}
\newpage
\section{Gated Architecture}\label{supp:gates}
In the past multiplicative interactions have been successful at stabilizing learning across different architectures \citep{multiplicativeInteraction, highway_networks}. This motivated us to try out GRU-type gating at the heads of the proposed Transformers. The outputs at the GRU-type gating is controlled by the following equation,
\begin{flalign*}
    r &= \sigma(W_r^{(l)}y + U_r^{(l)}x), \nonumber\\
    z &= \sigma(W_z^{(l)}y + U_z^{(l)}x - b_g^{(l)}), \nonumber\\
    \hat{h} &= tanh(W_g^{(l)}y + U_g^{(l)}(r \odot x)) \nonumber \\
    g^{(l)}(x,y) &= (1-z)\odot x + z \odot \hat{h} \nonumber \\
\end{flalign*}
where $r$ stands for the reset gates, $z$ is the update gates, $\hat{h}$ is the candidate activation similar to other recurrent units \citep{rnn}. The overall gate activation $g(x,y)$ takes input $x$ as the residual connection and $y$ the output of the \texttt{FeedForward} or \texttt{Multi-Head Attention} modules. $g(x,y)$ is basically an interpolation between the previous activations $\hat{h}$ and the residual input $x$.

\newpage
\onecolumn
\section{Speaking Rate for $\yo=0,1$}
\begin{figure*}[htbp!]
    \centering
    \includegraphics[width=\textwidth]{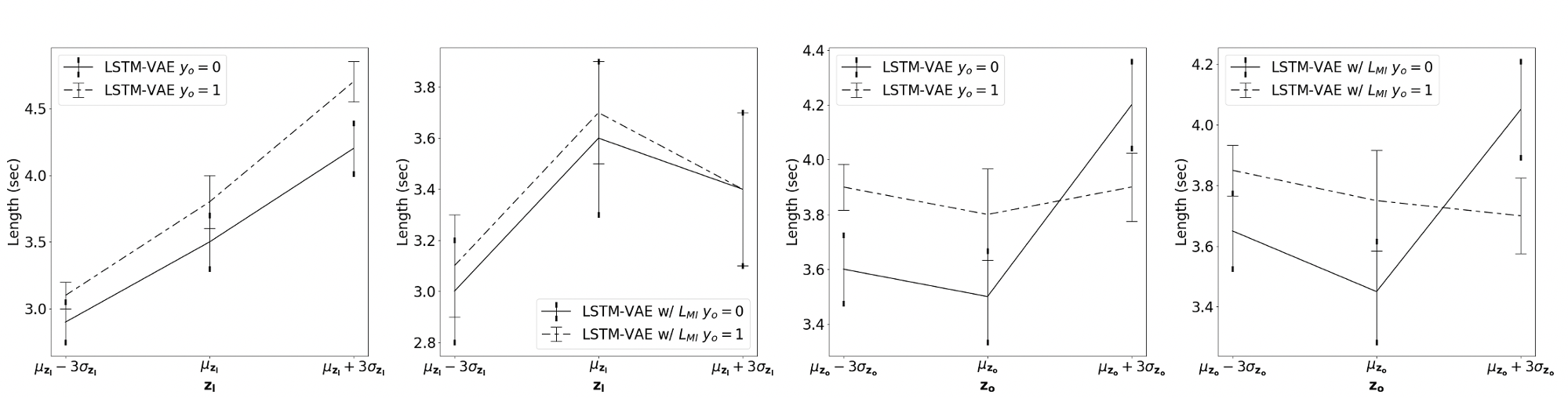}
\caption{Length of mel-spectrogram synthesized by $\zl$ in LSTM-VAE for MAILABS is significantly different for $\yo=0,1$ showing that $y_o$ specific information is encoded by $\zl$. However this difference is no longer significant once we include our proposed $L_{MI}$ terms in LSTM-VAE w/ $L_{MI}$ experiment. $\zo$ keeps showing different lengths for $y_o=0,1$ in both LSTM-VAE and LSTM-VAE w/ $L_{MI}$ experiments demonstrating learnt features which are conditional on $y_o$.
}
\label{fig:speaking_trends_lmi}
\end{figure*}{}

\begin{figure*}[htbp!]
    \centering
    \includegraphics[width=\textwidth]{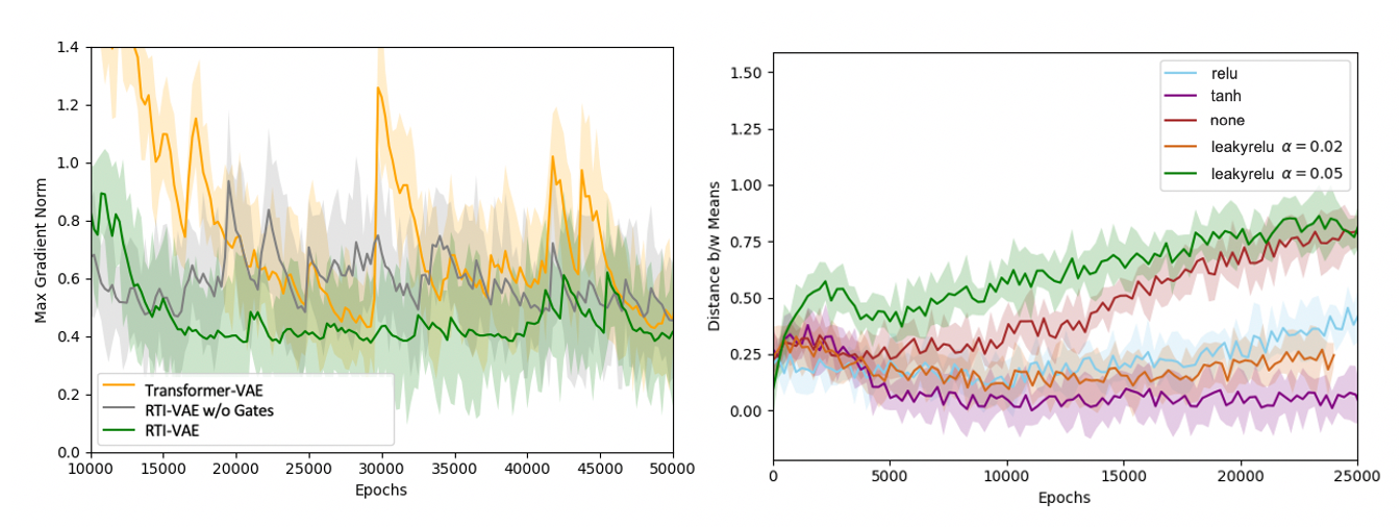}
\caption{
\textbf{Left}: Lower gradient norm for RTI-VAE w/ Gates along with smaller variance compared to Transformers-VAE and RTI-VAE w/o Gates.
\textbf{Right}: Distance between the means of $z_o|y_o$ for $y_o=0,1$ for different activation functions at the output of \texttt{Multi-Head Attention} and \texttt{FeedForward} modules. We see that $LeakyRelu$ with $\alpha=0.05$ performs the best in segregating the prior clusters among all experiments.
}
\label{fig:gradient_norm_activation_functions}
\end{figure*}{}

\twocolumn
\section{Ablation Study}
\subsection{Importance of Gates}\label{section:gradient_norm}
Our comparison of Gated architectures with non-Gated ones in Figure \ref{fig:gradient_norm_activation_functions} shows that the maximum gradient norm which directly influences the convergence is much lower and stable with a lower variance for RTI-VAE (which includes gates) compared to RTI-VAE without (w/o) Gates and Transformer-VAE.

\subsection{Choosing the Right Activation} \label{section:right_activation}
In Figure \ref{fig:gradient_norm_activation_functions} we see that the distance between $z_o|y_o$ cluster means is very small when the output from \texttt{Multi-Head Attention} and \texttt{FeedForward} modules are fed to \texttt{GRU-Type Gating} layers without any non linearity. Hence our choice of this non linearity was inspired by the trade-off between number of gradient updates and the maximum gradient norm. We see in Table \ref{tab:activation_table} that $relu$ has a high maximum gradient norm $\nabla_{norm}$ which led to convergence instability and small distance between $z_o|y_o$ cluster means. But for $tanh$, almost all activations were producing gradient updates and this frequent update was leading to small cluster distance as shown in Figure \ref{fig:gradient_norm_activation_functions}. Hence we needed an function somewhere between \texttt{relu} and \texttt{tanh}, which has a small gradient norm while also having fewer gradient updates. $LeakyRelu$ turns out to be the best candidate for this with its high distance between means as shown in Figure \ref{fig:gradient_norm_activation_functions}.

\begin{table}[H]
\centering
\begin{tabular}{l*{4}{c}r}
\hline \textbf{Experiment} & \textbf{\% activation}  & \textbf{max $\nabla_{norm}$}\\ \hline
    relu & 84.5 ($<0$) & 40.96 \\
    tanh & 0 ($>$+2,$<$-2) & 10.68 \\
    leakyrelu & - & 7.17 \\
\hline
\end{tabular}
\caption{Comparing the percentage of activations for which gradient saturates and maximum gradient norm $\nabla_{norm}$}
\label{tab:activation_table}
\end{table}

\section{Compute Information}
We ran all our experiments on NVIDIA Tesla V100 GPU with 16GB of GPU memory. Our LSTM-VAE (both with and without $L_{MI}$) experiments take average 5.81sec/step (seconds per step) with convergence near 40k steps. Transformer-VAE takes an average 2.81sec/step with convergence near 25k steps, and RTI-VAE takes average 2.81sec/step with convergence near 25k steps. Total number of parameters are 28.03mn (million) for LSTM-VAE w/ and w/o MI, 27.84mn for Tranformer-VAE and 28.03mn for RTI-VAE.

\section{Audio Hyperparameters}
\label{supp:audio_hparams}
\begin{table}[H]
    \label{tab:audio_params}
    \begin{center}
    \begin{tabular}{l*{2}{c}r}
    \toprule
    \textbf{Parameter}    & \textbf{Value}  \\
    \midrule
      num mels & 80\\
      num freq & 1025\\
      max mel frames & 900\\
      silence threshold & 2\\
      n fft & 2048\\
      hop size & 275\\
      win size & 1100\\
      sample rate & 16000\\
      magnitude power & 2.0\\
      trim silence & True\\
      trim fft size & 2048\\
      trim hop size & 512\\
      trim top db & 50\\
      preemphasize & True\\
      preemphasis & 0.97\\
      min level db & -100\\
      ref level db & 20\\
      fmin & 55\\
      fmax & 7600\\
      power & 1.5\\
    \bottomrule
    \end{tabular}
\end{center}
\caption{Parameters for converting wav files to mel-spectrograms}

\end{table}

\section{Tacotron-2 Hyperparameters}
\label{supp:tacotron_hparams}
\begin{table}[H]
\begin{center}
    \begin{tabular}{l*{2}{c}r}
    \toprule
    \textbf{Parameter}    & \textbf{Value}  \\
    \midrule
    batch size & 64 \\
    output frames per step & 4\\
    max training iterations & 100k\\
    optimizer & Adam \\
    \hspace{0.5cm} $\beta_1$ & 0.9 \\
    \hspace{0.5cm} $\beta_2$ & 0.999 \\
    \hspace{0.5cm} $\epsilon$ & 1e-6 \\
    L2 regularization weight & 1-e6 \\
    learning rate decay & exponential \\
    initial learning rate & 1e-3 \\
    decay start epoch & 40k \\
    decay epochs & 18k \\
    final learning rate & 1e-4 \\
    clip gradients & True \\
    teacher forcing & constant at 1 \\
    \bottomrule
    \end{tabular}
\end{center}
\caption{Hyperparameters common for all experiments}
\label{tab:common_ones}
\end{table}
\section{VAE Hyperparameters}
\label{supp:vae_hparams}

\begin{table}[H]
\begin{center}
    \begin{tabular}{l*{2}{c}r}
    \toprule
    \textbf{Parameter}    & \textbf{Value}  \\
    \midrule
    $z_l$ dim & 3 \\
    $z_o$ dim & 2 \\
    $|y_o|$ & 2 (UK, US) \\
    $z_o, z_l$ convolution channels & 128\\
    activation function for convolution & \texttt{tanh}\\
    kernel size & 3x3 \\
    MC estimate num\_samples & 1\\
    num\_units for LSTM & 128 \\
    min logvariance for $q(z_l|X)$ & -4 \\
    min logvariance for $q(z_o|X)$ & -6 \\
    initial mean for $p(z_l|y_l)$ & \\
    \hspace{0.5cm} $p(z_l|y_l=0)$ & (1,0,0) \\
    \hspace{0.5cm} $p(z_l|y_l=1)$ & (0,1,0) \\
    \hspace{0.5cm} $p(z_l|y_l=2)$ & (0,0,1) \\
    initial logvariance for $p(z_l|y_l)$ & -4 \\
    initial mean for $q(z_o|y_o)$ & \\
    \hspace{0.5cm} $p(z_o|y_o=0)$ & (-0.5, -0.5) \\
    \hspace{0.5cm} $p(z_o|y_o=1)$ & (+0.5, +0.5) \\
    initial logvariance for $p(z_o|y_o)$ & -5 \\
    dropout & 0.1 \\
    zoneout (for LSTM) & 0.1 \\
    $q_\psi$ num\_layers & 4\\
    $q_\psi$ num\_units & 8\\
    $q_\psi$ activations & \texttt{tanh}\\
    Transformer d\_model & 64 \\
    Transformer num\_heads & 4 \\
    Transformer feedforward\_dimension & 256 \\
    max positional encoding & 584\\
    \bottomrule
    \end{tabular}
\end{center}
\caption{Hyperparameters used for our VAEs}
\label{tab:latent_params}
\end{table}

\end{document}